\documentclass[12pt]{article}
\title{Transmutation of Elements through Capture of Electrons by Nuclei}
\author{A. Mukherji\footnote{Mailing address: D 111 Shanti Niket, 32/4
 Sahitya Parishad Street, Kolkata, 700006, India} \footnote{Electronic address:
 a.mukherji@saha.ac.in}  \\  \\
Formerly of Saha Institute of Nuclear Physics, \\
Block AF, BidhanNagar, Kolkata, 700064, India  }
\date{} \begin{document}
\maketitle
\begin{abstract}
A proton can capture an electron and turn into a neutron provided the electron 
has a kinetic energy of 0.782 MeV or more. An element of the Periodic Table
can change into another on being exposed to such high energy electrons.
\end{abstract}
\section{Introduction}
The possibility of converting Mercury into Gold intrigued scientists for ages.
The alchemists made considerable efforts towards this end without success.

Transmutation of one element into another takes place in radioactive 
decays and in nuclear reactions involving emission or absorption of charged
particles. Owing to poor yields these processes are not suitable for bulk
production of an element at a commercially viable cost.

Atoms split and unite in nuclear fission and fusion. Such priviledges are
enjoyed exclusively by a few members of the periodic table at its two
extremities.
\section{{\sl K}-Capture}
The nucleus of a heavy element can capture\cite{blatt} a {\sl K}-shell (1s) 
electron and thereby convert itself into a different element. This phenomenon 
is viewed as follows.

The attraction between a nucleus and its nearest neighbour 1s electron 
increases with increasing atomic number $Z$. The atomic orbital
shrinks in diameter. The electron itself comes closer to the nucleus
causing further increase in their mutual attraction. The net effect is
that the force between them varies as $Z^4$. When $Z$ is large the attraction 
becomes strong enough to release the electron from its bondage to its
orbital. It is then captured by the nucleus.

In this description the $1s$ electron is depicted as revolving around
the nucleus following a specified path and maintaining a definite
distance. This is, of course, far from being literally true. Every 
$s$-electron in an atom occasionally passes through the nucleus. Its
wavefunction is non-zero at the location of the nucleus. Hence, it has a
finite probability of being within the nuclear volume. This causes
hyperfine structures in atomic spectroscopy through Fermi\cite{slichter}
$\delta$-interaction. The nucleus only needs to hold on to a passing
electron. Of all the $s$-electrons the $1s$-electron is the most
energetic one. It travels the fastest through the nucleus. It is reasonable
to expect that the $1s$-electron will be the most difficult one to catch.
But in reality only $K$-shell electrons are captured by nuclei. The
slower $s$-electrons are ignored. The disparity between logical 
expectation and physical observation calls for a second look into the
matter.

The phenomenon of $K$-capture is discontinuous with respect to $Z$, while
the electron-nucleus force is a continuous function of $Z$. Explanation
of a discontinuous feature in terms of a continuous function is, in
principle, liable to be faulty.

The secondary cosmic rays that reach earth's surface are mostly slow
electrons. They are perpetually present in great abundance. They are free 
electrons and are not fettered to any orbital. They are easy prey for 
capture by any nucleus --- heavy or light --- on land or sea. No nucleus
has ever been known to have captured a cosmic ray electron. Otherwise, dire
consequences would have followed. Hydrogen nuclei everywhere on earth
would have turned into neutrons and the bound electrons would have gone free.
A great plasma of free neutrons and electrons would have resulted in
stead of living beings.
\section{Induced Transmutation}
When a nucleus captures an electron one of its protons turns into a neutron.
This change is permissible from the point of view of charge conservation
but not of mass conservation. The sum-total of the masses of a 
proton (938.272 MeV) and an electron (0.511 MeV) falls short of the mass of 
a neutron (939.565 MeV) by 0.782 MeV. Hence proton to neutron conversion
is prohibited unless the captured electron provides this extra mass in the
form of kinetic energy. It is very large for an electron with rest mass of 
about 0.5 MeV and is in the relativistic range. No nucleus can, therefore,
capture a slow electron. Only $1s$-electrons in very heavy atoms can
provide the required energy. Hence, $K$-capture is limited to them only.
The minimum energy requirement is, however, subject to a small correction
due to changes in binding energies of the nucleus and of the electrons in
various orbitals associated with the change in the constitution of the
nucleus.

Transmutation of an element $X$ into another $Y$ can be induced by
irradiating $X$ by a strong beam of high energy electrons of at least
0.782 MeV. The new element $Y$ can be extracted from the exposed target by
chemical means. The process is complete if the isotope of $Y$ so
produced is a stable one. Otherwise $Y$ may undergo radioactive decay and
get converted into some other element.

It is now a matter of technology and experimentation to produce any desired 
element with high yield at low cost. The output will be chemically
indistinguishable from that obtained from natural sources. The difference
will only be in their isotopic constituents.

If large quantities of gold and/or other noble metals could be produced 
cheaply, they would loose their elevated positions as asset materials. 
World economy will suffer a jolt and it will be in need of a new
standard.  


\begin{thebibliography}{99}
\bibitem{blatt} John M Blatt and Victor F Weisskopf, {\it Theoretical
  Nuclear Physics}, (John Wiley, New York, 1952)
\bibitem{slichter} C.P.Slichter, {\it Principles of Magetic Resonance},
  (Springer, Berlin, 1980) 
\end{thebibliography}
\end{document}